%% file: main.tex
\documentclass[twocolumn]{aastex702}

\makeatletter

\let\original@olddoauthor\olddoauthor

\def\olddoauthor#1#2#3{%
  \original@olddoauthor{#1}{#2}{#3}%
  \ifnum\countauthors=4\relax
    \unskip\break
  \fi
}

\makeatother

\newcommand{\be}{\begin{displaymath}}
\newcommand{\ee}{\end{displaymath}}
\newcommand{\bea}{\begin{eqnarray*}}
\newcommand{\eea}{\end{eqnarray*}}

\shorttitle{Perpendicular Orbits of HD\,3167?}
\shortauthors{Winn et al.}

\graphicspath{{./}{}}
\usepackage{amsmath}
\usepackage{comment}

\usepackage{hyperref}
\hypersetup{
    colorlinks=true,
    linkcolor=blue,
    filecolor=magenta,      
    urlcolor=cyan,
    pdftitle={Overleaf Example},
    pdfpagemode=FullScreen,
    }

\begin{document}

\def\ltsima{$\; \buildrel < \over \sim \;$}
\def\lsim{\lower.5ex\hbox{\ltsima}}
\def\gtsima{$\; \buildrel > \over \sim \;$}
\def\gsim{\lower.5ex\hbox{\gtsima}}

\title{Does HD\,3167 Have Planets with Perpendicular Orbits?}

\correspondingauthor{Joshua N.\ Winn}
\email{jnwinn@princeton.edu}

\author[0000-0002-4265-047X]{Joshua N.\ Winn}
\affiliation{Department of Astrophysical Sciences, Princeton University, 4 Ivy Lane, Princeton, NJ 08544, USA}
\email{jnwinn@princeton.edu}

\author[0000-0001-8342-7736]{Jack Lubin}
\affiliation{Department of Physics \& Astronomy, University of California Los Angeles, Los Angeles, CA 90095, USA}
\email{}

\author[0000-0001-7409-5688]{Guðmundur Stefánsson}
\affiliation{Department of Astrophysical Sciences, Princeton University, 4 Ivy Lane, Princeton, NJ 08544, USA}
\affil{Anton Pannekoek Institute for Astronomy, 904 Science Park, University of Amsterdam, Amsterdam, 1098 XH}
\affil{Astrophysics \& Space Center, Schmidt Sciences, New York, NY 10011, USA}
\email{}

\author[0000-0002-0971-6078]{Haochuan Yu}
\affiliation{Department of Physics,
University of Oxford, Oxford OX1 3RH, UK}
\affiliation{Department of Astrophysical Sciences, Princeton University, 4 Ivy Lane, Princeton, NJ 08544, USA}
\email{}

\author[0000-0003-0967-2893]{Erik Petigura}
\affiliation{Department of Physics \& Astronomy, University of California Los Angeles, Los Angeles, CA 90095, USA}
\email{}

\author[0000-0002-0531-1073]{Howard Isaacson}
\affiliation{Department of Astronomy, University of California at Berkeley, Berkeley, CA 94720, USA}
\email{}

\author[0000-0001-8638-0320]{Andrew W.\ Howard}
\affiliation{Department of Astronomy, California Institute of Technology, Pasadena, CA, USA}
\email{}

\author[0000-0002-8958-0683]{Fei Dai}
\affiliation{Institute for Astronomy, University of Hawaii, 2680 Woodlawn Drive, Honolulu, HI 96822 USA}
\email{}

\begin{abstract}

The two transiting planets of HD\,3167 were reported to follow
nearly perpendicular paths, with the inner 
orbit aligned with
the stellar equator and the outer orbit nearly polar.
This interpretation depends critically on a
challenging single-transit detection of the
Rossiter-McLaughlin effect of the inner planet.
We observed three additional transits of the inner planet with the Keck Planet Finder and analyzed the new data together with two archival ESPRESSO transit datasets.
We do not confirm the previously
reported low obliquity.
Our analysis favors a projected
obliquity of $-66^{+14}_{-12}$~deg,
consistent with coplanar orbits.
However, because the best-fit 
projected rotation velocity
is higher than expected, and because the
obliquity uncertainty grows substantially
when the most discrepant of the five transit datasets is omitted,
we regard the geometry
of the HD\,3167 system as still unsettled.
\end{abstract}

\section{Introduction}

Sun-like stars often have
several planets with periods of days to months and
sizes between those of Earth and Neptune \citep{Lissauer+2011}. Several lines of evidence
indicate that such systems are usually flat.
When multiple planets are transiting, the observed trend of transit durations with orbital distance implies mutual inclinations of only a few degrees \citep{Fabrycky+2014}.
Analyses of transit-timing variations have also implied
coplanar orbits in many compact systems
\citep[e.g.,][]{SanchisOjeda+2012, Zhu+2018}, as do
comparisons between transit and radial-velocity survey results
\citep{TremaineDong2012}. Even among the known exceptions,
the reported mutual inclinations are usually modest,
$\lesssim 25^\circ$ \citep[e.g.,][]{MillsFabrycky2017, Almenara+2022}.

Against this backdrop, the
HD\,3167 system stands out.
It has two transiting planets within 0.2~AU
\citep{Vanderburg+2016, Gandolfi+2017, Christiansen+2017}, both of which
were subject to Rossiter-McLaughlin (RM)
observations to assess the alignment
between the orbit and stellar spin.
\cite{Dalal+2019} reported a polar
orbit for the outer planet,
while \cite{Bourrier+2021}
found a nearly equatorial
orbit for the inner planet.
Taken together, these measurements imply that the
planets have nearly perpendicular orbits.
Such a configuration would be extraordinary
among compact multiplanet systems,
with implications for theories of planet formation and dynamical evolution.
Furthermore, because two perpendicular orbits are unlikely
to be viewed edge-on from the same line of sight
\citep{Teng+2025},
even one such system might imply a large
population of strongly misaligned systems.

The RM measurement was especially challenging for the
inner planet because the expected radial velocity
anomaly is less than a meter per second.
The combination of a striking proposed architecture
and a difficult measurement 
prompted us to collect more data.
We observed three additional transits of the inner planet,
and analyzed the data alongside two archival observations. 

Section~\ref{sec:review} summarizes the HD\,3167 system and the previous obliquity measurements. Section~\ref{sec:observations} describes the observations. Section~\ref{sec:vsini} discusses the constraints on the star's
rotation velocity, which are important for the interpretation of the data.
Section~\ref{sec:rm} presents our analysis of the RM effect, and Section~\ref{sec:discussion} discusses the implications for the system's orbital architecture.

\section{The HD\,3167 system}
\label{sec:review}

HD\,3167 is a chromospherically
quiet K0 dwarf at a distance of 46 pc, with $V=8.9$.
The two transiting planets were discovered by \cite{Vanderburg+2016}.
Planet b is a super-Earth ($1.7\,R_\oplus$) with an orbital period of about
one day, and planet c is a sub-Neptune ($3.0\,R_\oplus$) with a 30-day period.
Follow-up radial-velocity observations by
\cite{Gandolfi+2017} and \cite{Christiansen+2017} established
their masses to be $M_b\approx 5\,M_\oplus$ and $M_c\approx 9\,M_\oplus$.
\cite{Christiansen+2017} also reported evidence for a non-transiting planet, d,
with a minimum mass of 6.9~$M_\oplus$ and a period of 8.5 days.
A broader system analysis by \cite{Bourrier+2022}
revised d's minimum mass to 5.0~$M_\oplus$
and found tentative evidence for another non-transiting planet, e,
with a minimum mass of 9.7~$M_\oplus$ and a
period between about 79 and 125 days.

The sky-projected obliquity of planet c
was measured
by \cite{Dalal+2019} based on RM observations
with the HARPS-N spectrograph.
After considering several approaches to modeling the RM effect,
they concluded
$\lambda_c = -97^\circ \pm 23^\circ$, indicating a nearly polar orbit.
\cite{Bourrier+2021}
subsequently observed a transit of planet b with ESPRESSO.
They employed an analysis method referred to
as ``RM revolutions'' involving a model for the planet-induced
distortions in the
cross-correlation functions (CCFs).
Applied to the HARPS-N data for planet c, \cite{Bourrier+2021}
determined $\lambda_c = -108.9^{+5.4}_{-5.5}$~deg.
Incorporating the ESPRESSO data for planet b,
they found $\lambda_b = -6.6^{+6.6}_{-7.9}$~deg.
This combination is the basis of the
proposed perpendicular arrangement.

\section{Observations and data reduction}
\label{sec:observations}

We observed three transits of HD\,3167\,b
with the Keck Planet Finder (KPF) on the Keck I telescope on
Mauna Kea, Hawaii.
KPF is a stabilized, fiber-fed optical
echelle spectrograph with separate green and red channels,
covering approximately 445--600~nm and 600--870~nm, respectively,
at a resolving power of $R\approx 98{,}000$ \citep{Gibson+2024}.
The observations were conducted on UT~2023~September~22, September~23, and October~20.
They were designed to span full transits,
based on the ephemeris of \cite{Bourrier+2022}.
The exposure time was 300~sec, resulting
in 50, 53, and 54 spectra on the three nights.
Conditions were mostly clear and observing progressed smoothly
apart from a 30~min interruption on the first night due to a temporary
failure of the tip/tilt system.
The spectra were processed with version 2.7.3 of
the KPF Data Reduction Pipeline. We used the radial velocities
from the green channel (header keyword {\tt CCD1RV}),
for which the median radial-velocity uncertainty is 0.6~m/s.
We did not use the red-channel velocities because they
have significantly higher uncertainties.

We also analyzed two archival ESPRESSO transit observations of HD\,3167\,b.
ESPRESSO is a stabilized, fiber-fed optical
echelle spectrograph that can be connected
to any of the Very Large Telescopes at Paranal Observatory, Chile.
The high-resolution mode was employed,
with resolving power $R\approx 140{,}000$ \citep{Pepe+2021}. 
The first observation was on UT~2019~October 9
with VLT\,UT3, 
and was the basis of the low-obliquity measurement
by \cite{Bourrier+2021}.\footnote{Although 
\cite{Bourrier+2021} reported the date of this observation
as 2019~February~20, the FITS headers of the data files
specify 2019~October~9.}
The second observation
was conducted under the auspices
of the ESPRESSO Guaranteed Time Observation program
(P.I.\ F.\ Pepe)
on UT~2022~November~3 with UT2 and, to our knowledge,
has not previously been described in the literature.
Radial velocities were extracted from the ESPRESSO cross-correlation functions
(keyword {\tt HIERARCH ESO QC CCF RV}).
The 2019 dataset consists of 39 radial velocities from 300~sec exposures,
with a median uncertainty of 0.8~m/s.
The 2022 dataset comprises 58 radial velocities
based on 200~s exposures,
with a median uncertainty of 0.5~m/s. The lower uncertainty
may be due, at least in part, to better seeing ($0.7''$ in 2022
versus $2''$ in 2019).
Table~\ref{tbl:rv_kpf_espresso} gives the radial velocity
measurements.

\input{rv_kpf_espresso_abbr.tex}

\section{Stellar rotation and line width}
\label{sec:vsini}

\begin{figure*}
\centering
\includegraphics[width=1.0\textwidth]{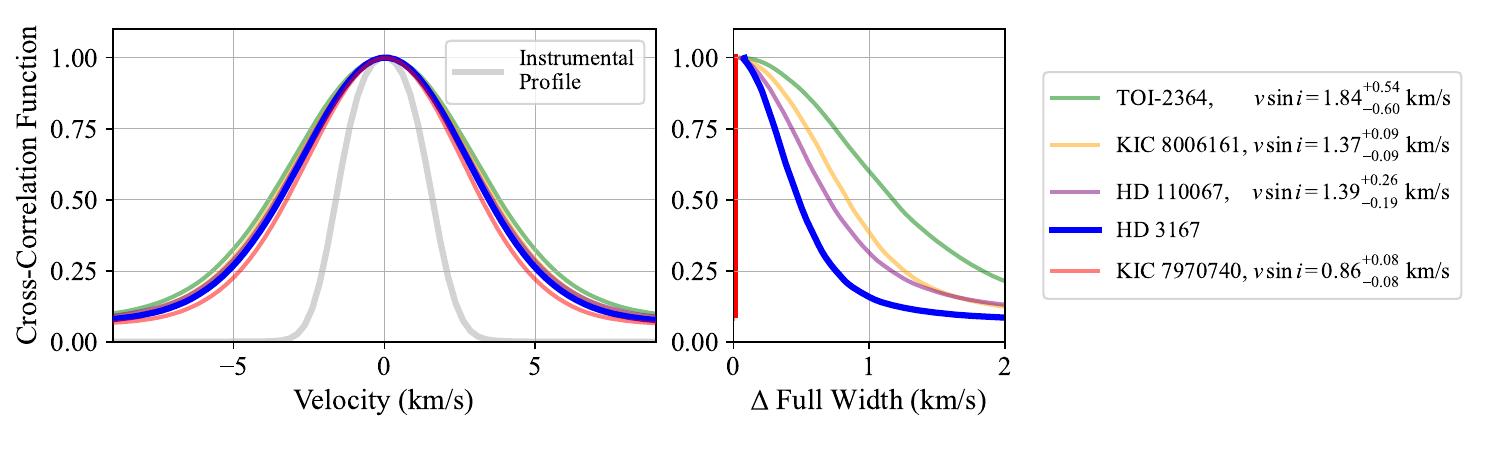}
\vskip -0.2in
\caption{KPF cross-correlation functions (CCFs)
for HD\,3167 and four comparison stars with similar atmospheric parameters and independently measured
projected rotation velocities.
(Left) Normalized CCFs
and the KPF instrumental profile.
(Right) Differences between each star's full width and
the full width of the narrowest profile (KIC\,7970740), evaluated at each normalized CCF level.
HD\,3167 has a narrower profile
than the stars
with $v\sin i \approx 1.4$~km/s and broader than
the star with $v\sin i \approx 0.9$~km/s,
suggesting its projected rotation velocity
is between those values.}
\label{fig:comparison_stars}
\end{figure*}

The amplitude of the RM signal depends not
only on the projected obliquity $\lambda$,
but also on the projected stellar rotation velocity, $v\sin i_\star$ \citep{GaudiWinn2007}.
External constraints on $v\sin i_\star$ are often useful
when fitting RM data, especially when
the signal is near the margin of detectability,
as is the case for HD\,3167\,b.

The published determinations of $v\sin i$ for HD\,3167 based on spectral-line broadening
are $<$2~km~s$^{-1}$ \citep{Vanderburg+2016},
$1.8\pm0.4$~km~s$^{-1}$ \citep{Gandolfi+2017},
and $1.7\pm1.1$~km~s$^{-1}$ \citep{Christiansen+2017}.
Such measurements are
inherently challenging
for slowly-rotating stars because
rotational broadening is comparable to 
macroturbulent broadening,
and both effects are often comparable to or smaller than the
instrumental resolution (\citealt{Doyle+2014}, Lubin et al.\ submitted).
In principle, the
broadening mechanisms
can be distinguished because their
kernels have different shapes.
In practice, the distinction requires exceptionally precise spectra and hinges
on the accuracy of the assumed line-profile models.

For a test that is less dependent on line-profile modeling,
we compared the KPF CCF of HD\,3167 with those of four
stars chosen to have similar atmospheric parameters
and for which independent measurements of
$v\sin i$ are available
(Table~\ref{tbl:comparison_stars}):
from the RM effect, for HD\,110067 and TOI-2364
\citep{Zak+2024, Tamburo+2025}, and from asteroseismology,
for KIC\,7970740 and KIC\,8006161
(\citealt{Kamiaka+2018, Hall+2021}; Lubin et al., submitted).

Figure~\ref{fig:comparison_stars} shows that the
CCF width of HD\,3167 is broader
than that of KIC\,7970740, which has
$v\sin i = 0.86\pm 0.08$~km/s.
It is also narrower than the CCF widths
of HD\,110067 and KIC\,8006161,
which both have $v\sin i \approx 1.4$~km/s.
Since instrumental broadening is the same in all cases,
and macroturbulent broadening
is likely similar\footnote{The last row of
Table~\ref{tbl:comparison_stars}
gives the macroturbulent velocity
for each star
predicted by the empirical
relations of \cite{Doyle+2014}, based on
effective temperature and surface gravity.},
this comparison
suggests that the $v\sin i$ of HD\,3167 is
between about 0.9 and 1.4~km/s,
and is unlikely to be greater than 1.6~km/s.

Another handle on $v\sin i$ comes from estimates
of the star's radius and rotation period.
Using the light curve from the K2 mission, 
\cite{Christiansen+2017} reported a broad autocorrelation
peak near 27 days, \cite{Gandolfi+2017} found
a period of $23.5\pm2.9$~days,
and \cite{Bourrier+2022} refined this
estimate to $23.4\pm2.2$~days.
\cite{Bourrier+2022} also
estimated a rotation period of
$24.1\pm1.2$~days from the periodogram of radial velocities
and spectroscopic activity indicators.
For $R_\star \approx 0.87\,R_\odot$, a rotation
period of 24~days
corresponds to an equatorial rotation speed
of about 1.8~km/s.

Taken together, the line-width and rotation-period evidence
point to $v\sin i_\star \lesssim 1.8$~km/s.
This is in mild tension with
the value inferred in the RM-based analysis of \cite{Bourrier+2021},
$v\sin i_\star=2.41\pm 0.37$~km~s$^{-1}$,
which corresponds to a maximum equatorial rotation period
of about 18 days. We return to discuss this
tension in Section~\ref{sec:discussion}.

\begin{deluxetable*}{lccccc}
\label{tbl:comparison_stars}
\tablecaption{Key properties of HD\,3167 and four comparison stars.}
\tablehead{
Parameter & HD\,3167 & KIC\,7970740 & HD\,110067 & KIC 8006161 & TOI-2364 
}
\startdata
Effective temperature (K) & 
    $5300\pm 73$ (1) &
    $5309\pm 77$ (2) & 
    $5266\pm 64$ (3) &
    $5488\pm 77$ (2) &
    $5306_{-68}^{+76}$ (5) 
    \\
$\log g$ (cgs) &
    $4.47\pm 0.12$ (1) &
    $4.54\pm 0.01$ (2) &
    $4.54\pm 0.03$ (3) &
    $4.49\pm 0.01$ (2) &
    $4.524_{-0.029}^{+0.018}$ (5)
    \\
$[{\rm Fe}/{\rm H}]$ &
    $+0.04 \pm 0.05$ (1) &
    $-0.49 \pm 0.10$ (2) &
    $-0.20 \pm 0.04$ (3) &
    $+0.34 \pm 0.10$ (2) &
    $0.34 \pm 0.09$ (5)
    \\
$v\sin i$ (km/s) &
    $\cdots$ & 
    $0.86\pm 0.08$ (2) &
    $1.39^{+0.26}_{-0.19}$ (4) & 
    $1.37\pm 0.09$ (2) &
    $1.840_{-0.060}^{+0.054}$ (6)
    \\
Predicted $v_{\rm mac}$ (km/s) &
    $2.51\pm 0.24$ &
    $2.37\pm 0.05$ &
    $2.35\pm 0.06$ &
    $2.61\pm 0.09$ &
    $2.40 \pm 0.07$
\enddata
\tablecomments{
The $v_{\rm mac}$ predictions are based on Eq.~8 of \cite{Doyle+2014},
with uncertainties reflecting propagated uncertainties in $T_{\rm eff}$ and $\log g$.
References: 
(1) \cite{Bourrier+2022};
(2) \cite{Hall+2021};
(3) \cite{Luque+2023};
(4) \cite{Zak+2024};
(5) \cite{Yee+2023};
(6) \cite{Tamburo+2025}.
}
\end{deluxetable*}

\section{Rossiter-McLaughlin analysis}
\label{sec:rm}

We analyzed the RM effect by modeling the
time series of apparent radial velocities. 
For a given radial velocity measurement $v_i$ obtained at time $t_i$
within the $j$th of the five transit sequences,
the model is
\[
v_i = v_{\rm orb}(t_i) + \Delta v_{\rm RM}(t_i) + 
     \gamma_j + \dot{\gamma}_j (t_i-t_{c,j}).
\]
Here, $v_{\rm orb}$ is the star's reflex motion due to planet b,
\[
v_{\rm orb}(t_i) = -K\sin\left[ \frac{2\pi}{P} (t-t_{c,j}) \right],
\]
where $P$ is the orbital period, $K$ is the 
velocity semi-amplitude, and $t_{c,j}$ is the time of conjunction.
The anomalous velocity $\Delta v_{\rm RM}$ depends on
$v\sin i$, $\lambda$, and the geometric
transit parameters. We used the analytic prescription of
\cite{Hirano+2010}, with the occulted flux
computed with the analytic transit model of
\citet{MandelAgol2002}.\footnote{We adopted a linear limb darkening law with coefficient 0.6, and a non-rotational broadening parameter $\beta=2.6$~km/s,
but varying these parameters over reasonable ranges
does not materially change the results.}
Each transit observation had its own
velocity offset $\gamma_j$ and linear trend
$\dot{\gamma}_j$, to account
for the arbitrary velocity zero point of each
instrument as well as slow trends due to
other planets, stellar activity, or instrumental drift.
Since $\dot{\gamma}$ is partially degenerate with $K$,
we fixed $K = 3.425$~m/s \citep{Bourrier+2022} and allowed
the uncertainty in this value
to be absorbed by the linear trends.
For the noise model,
we assumed the RVs have
independent Gaussian errors with variances
equal to the uncertainties reported
by the KPF and ESPRESSO data reduction pipelines; we did not find it necessary
to allow for excess ``RV jitter.''

Table~\ref{tbl:photometric-parameters} gives
the adopted priors on the ephemeris and transit parameters,
drawn from \cite{Bourrier+2022}.
Broad uniform priors were adopted for the other parameters.
Parameter optimization was performed with the
Levenberg-Marquardt code {\tt LMFIT} 
\citep{Newville+2014},
and samples of the posterior probability distribution
were generated with the Markov Chain Monte Carlo
code {\tt emcee}
\citep{emcee}.\footnote{
Each of 96 walkers took $10^5$ steps, of which the initial
20\% were discarded.
We assessed convergence by inspecting the walker traces,
comparing posterior summaries from the first and second halves of the chains,
and estimating autocorrelation lengths.
The retained chain length per walker was much longer
than the estimated autocorrelation time for all parameters.}

\begin{deluxetable}{lc}
\label{tbl:photometric-parameters}
\tablecaption{Priors on HD\,3167\,b parameters.\label{tbl:priors}}
\tabletypesize{\small}
\tablewidth{0pt}
\tablehead{
Parameter & Value  
}
\startdata
Time of conjunction, $T_{\rm c,0}$ (BJD) & 2458269.57891(66) \\
Orbital period, $P$ (days)                         & 0.95965428(30) \\
Radius ratio, $R_{\rm p}/R_\star$                  & 0.01712(90) \\
Scaled orbital distance, $a/R_\star$               & 4.450(55) \\
Total transit duration, $T_{14}$~(hr)              & 1.609(17) 
\enddata
\tablecomments{Numbers in parentheses are the uncertainties
in the last two digits. Taken from \cite{Bourrier+2022}, with uncertainties
rounded up in some cases.
The parameter $T_{14}$ was used
rather than the impact parameter $b$
in order to reduce parameter covariance.
}
\end{deluxetable}

\begin{figure*}
\includegraphics[width=1.0\textwidth]{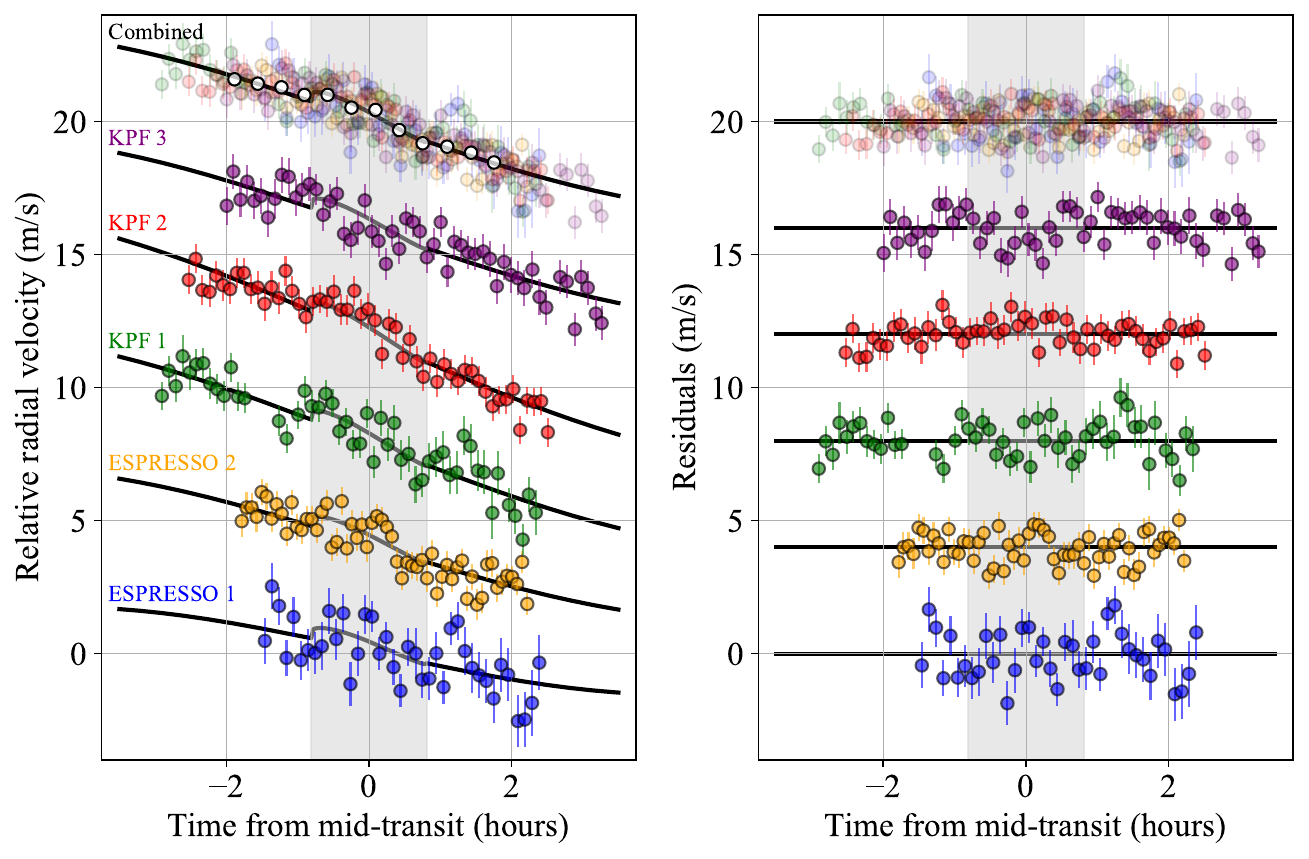}
\caption{
The left panel shows the apparent radial velocities of HD\,3167\,b
spanning five separate transits (colored points)
and the best-fit consensus model
including the RM effect (black curves).
The right panel shows the residuals after subtracting the best-fit model.
Arbitrary vertical offsets were applied
to separate data from different nights.
The lowermost dataset was
analyzed by \cite{Bourrier+2021}.
The uppermost is a composite of all the data,
including phase-binned points for display purposes.
The gray region is the transit interval.
}
\label{fig:HD3167_RM_fits}
\end{figure*}

\begin{figure*}
\centering
\includegraphics[width=0.9\textwidth]{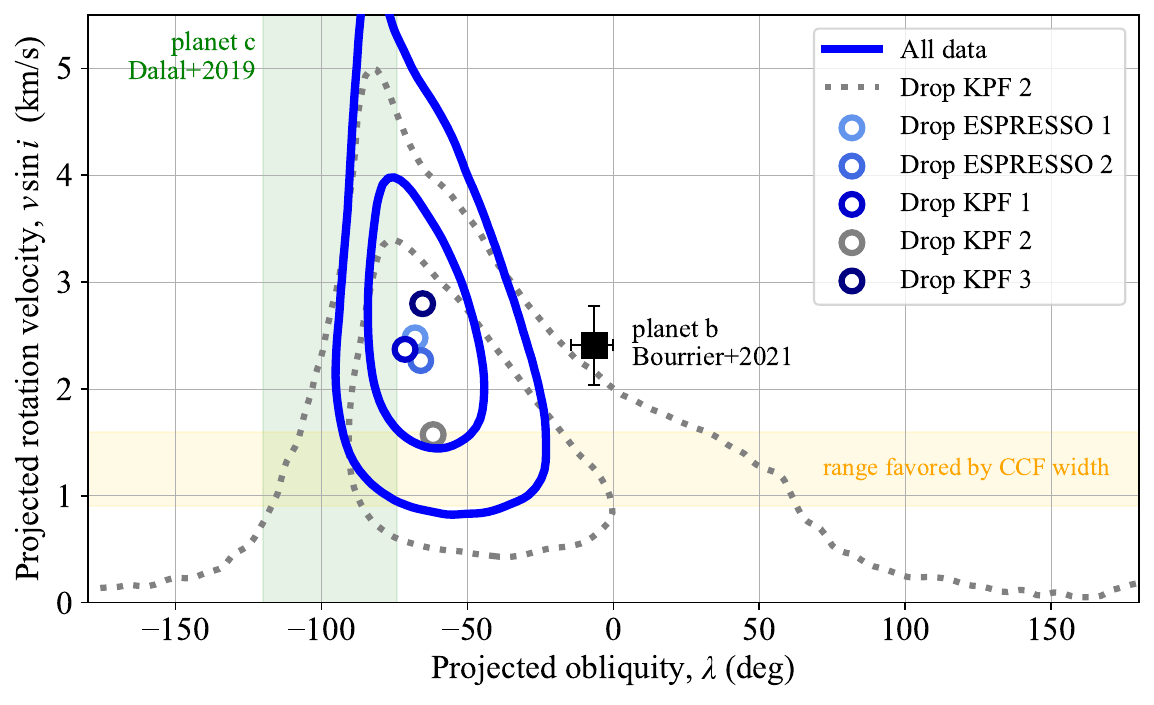}
\caption{Posterior probability density for
the $v\sin i$ and $\lambda$.
Blue contours are the 68\% and 95\% levels
based on fitting all 5 datasets.
Dotted gray contours show the results of dropping
the second KPF observation, the most
discrepant of the five datasets.
The circular data points show the maximum
{\it a posteriori} results from each of
the leave-one-out tests.
The black square is from \cite{Bourrier+2021}.
The yellow zone illustrates the range 
$v\sin i \approx 0.9$--1.6~km/s
favored by the CCF comparisons presented
in Section~\ref{sec:vsini}.
The green zone shows the measured
projected obliquity of planet c \citep{Dalal+2019}.
}
\label{fig:HD3167_MCMC}
\end{figure*}

Figure~\ref{fig:HD3167_RM_fits} displays the
data and the best-fit consensus model, which has
$\chi^2=275.3$ with 247 degrees of freedom ($p=0.1$).
Figure~\ref{fig:HD3167_MCMC} shows
the posterior probability density
in the $\lambda$--$v\sin i$ plane.
The posterior medians
and central 68\% credible intervals are
\[
\lambda_b = -66^{+14}_{-12}~{\rm deg},~~
v\sin i = 2.54^{+0.93}_{-0.73}~{\rm km/s}.
\]
This solution is far from the low-obliquity
result reported by \citet{Bourrier+2021}.

To assess the statistical significance of
the detection,
we fitted a null model that 
allows for nightly slopes
and offsets but contains no RM effect.
The best-fit null model
has $\chi^2=296.5$ with 249 degrees of freedom ($p=0.02$). The RM model
is preferred over the null model
by $\Delta\chi^2 = 21.2$ and 
$\Delta$BIC (Bayesian Information Criterion) of 10.1.

We also checked
whether our analysis procedure
could have detected a signal with
the characteristics of the previously
reported signal. We created simulated
RVs based on a model with the
\cite{Bourrier+2021}
parameters, using the
same timestamps as the actual RVs
and adding Gaussian perturbations
according to the actual RV uncertainties.
Fitting the simulated data with the same
code as the actual data gave
$\lambda_b = 9\pm 17$~deg
and $v\sin i = 2.9\pm 0.6$~km/s,
demonstrating that our procedure
could have recovered such a signal.

To check on whether our results
are driven by any single
transit dataset, we repeated the optimization five times,
each time omitting one of the five datasets.
The circular data points in Figure~\ref{fig:HD3167_MCMC}
show the resulting maximum {\it a posteriori} values.
The results for $\lambda_b$ agree
with each other, as expected if the signal is present in all datasets.
The inferred $v\sin i$, however, drops by about 1-$\sigma$
when the second KPF dataset is excluded.
For this reason,
Figure~\ref{fig:HD3167_MCMC} also shows the
posterior density after excluding that dataset.
In that case, the preferred $v\sin i$ 
is lower and more compatible
with the constraints on $v\sin i$
described in Section~\ref{sec:vsini},
and the allowed
range of $\lambda_b$ is broader.

The black square in Figure~\ref{fig:HD3167_MCMC}
marks the previous result of \cite{Bourrier+2021}
for planet b.
The green band 
illustrates the measurement by
\cite{Dalal+2019}
of the projected obliquity of planet c.
The large separation between these two
measurements is the basis
of the inference that their orbits
are nearly perpendicular.
In contrast, the posterior density from
our anomalous-RV analysis lies
far from the low-obliquity
solution for planet b,
and overlaps the projected-obliquity
range measured for planet c.

\section{Discussion}
\label{sec:discussion}

The purpose of this work was to test whether the previously reported
low obliquity of HD\,3167\,b can be
recovered from an expanded dataset
using a simple model for the RM effect.
We found that it cannot.
The full dataset favors a large projected
misalignment for planet b, closer to the measurement
for planet c than the low-obliquity solution
reported by \citet{Bourrier+2021}.
This result is compatible
with coplanar orbits.

Although our analysis does not support
the prior claim of perpendicular orbits,
neither does it rule out a large mutual inclination.
Using the transit constraints on the inclinations together with our result
for $\lambda_b$, and
$\lambda_c=-97\pm23^\circ$ from \cite{Dalal+2019},
the 90\%-confidence upper limit
on the three-dimensional
mutual inclination is $66^\circ$, a weak constraint.

Furthermore,
there are reasons for caution in interpreting the formal parameter posteriors.
Using the full dataset,
the inferred value of $v\sin i$
is 1-$\sigma$ higher than expected from the constraints on the stellar
line width and rotation period discussed in Section~\ref{sec:vsini}.
This tension is reduced when the most discrepant of the five datasets
is omitted, but 
in that case, the constraints
on $\lambda$ are looser.
For example, assuming $v\sin i = 1$~km/s, the data
allow $\lambda_b$ to have any value
from $-25^\circ$ to $-75^\circ$ (Figure~\ref{fig:HD3167_MCMC}).
Thus, while our analysis does not support the claim
of perpendicular orbits, it also does not provide a
robust replacement measurement.

The previous result was based on one-fifth
of the current data and yet led to tighter constraints
on $\lambda$ and $v\sin i$.
The key difference is probably in the analysis methods.
Rather than reducing each spectrum to an apparent radial velocity,
\citet{Bourrier+2021} modeled the time series of CCFs.
This approach has the advantage, in principle, of harvesting
more information than the centroid shift alone.
For HD\,3167\,b, however, the situation is not ideal for this
approach. The planet is small, the star is a slow rotator,
and the expected line-profile
distortions are 
too weak to be detected in individual CCFs.
In this regime, the inferred obliquity
may depend critically on assumptions about the intrinsic
stellar line profile and its variation across the stellar disk.

Several features of the previously reported model illustrate
this sensitivity.
The inferred $v\sin i$ is higher than suggested by the line-width
and rotation-period constraints, a discrepancy 
attributed to strong differential rotation.
In addition, the joint analysis of data for both planets
involved an unexpected latitude-dependent
local line contrast. These effects are physically possible,
but the signals were not well resolved or strong enough
to validate those ingredients independently.
A line-profile analysis of the expanded dataset
would be valuable if it could help to test whether
the same ingredients are demanded by the new data.

For now, the most conservative conclusion is that the
mutual inclination between the two transiting planets
in the HD\,3167 system cannot be robustly determined
with the available data.
From this stance, coplanar orbits are a reasonable default assumption,
both because they are typical of compact multiplanet systems
and because observing transits of both planets
requires a more finely tuned viewing direction
when the orbits are perpendicular than when
they are aligned \citep{BrakensiekRagozzine2016, Teng+2025}.
Finally, our work suggests that
if we are ever to find
transiting planets with 
perpendicular orbits,
we should hope they revolve around rapidly
rotating stars.

\section{Acknowledgments} 
\label{sec:acknowledgments}

We thank V.~Bourrier and the anonymous referee for 
helpful comments on the manuscript.
We are grateful to Nick Saunders,
Aaron Householder,
Steven Giacalone,
Corey Beard,
and Jacob Luhn
for conducting the KPF
observations remotely.
This work was supported by a NASA Keck PI Data Award, administered by
the NASA Exoplanet Science Institute.
Data presented herein were
obtained at the W.\ M.\ Keck Observatory from telescope time allocated
to the National Aeronautics and Space Administration through the
agency's scientific partnership with the California Institute of
Technology and the University of California. The Observatory was made
possible by the generous financial support of the W.\ M.\ Keck
Foundation.
The authors wish to recognize and acknowledge the very
significant cultural role and reverence that the summit of Mauna Kea
has always had within the indigenous Hawaiian community. We are most
fortunate to have the opportunity to conduct observations from this
mountain.

\facility{Keck:I, VLT}

\bibliography{refs}{}
\bibliographystyle{aasjournal}

\end{document}

%% file: rv_kpf_espresso_abbr.tex
\begin{deluxetable}{@{}rccc@{}}
\tablecaption{Relative RV measurements of HD\,3167.
\label{tbl:rv_kpf_espresso}}
\tablewidth{0pt}
\tabletypesize{\small}
\tablehead{
\colhead{BJD} &
\colhead{RV\,(m/s)} &
\colhead{$\sigma_{\rm RV}$\,(m/s)} &
\colhead{Instrument}
}
\startdata
2460209.87884 & 1.685 & 0.531 & KPF \\
2460209.88286 & 2.633 & 0.566 & KPF \\
2459886.68594 & -0.553 & 0.414 & ESPRESSO \\
2459886.68891 & -2.132 & 0.419 & ESPRESSO \\
\enddata
\tablecomments{This is an abbreviated table
illustrating its format; the entire table
will be published in the electronic journal.}
\end{deluxetable}